\documentclass[epjST]{svjour}
\usepackage{graphicx}
\usepackage{graphicx,epstopdf}
\usepackage{graphics}
\begin{document}
\title{Synchronization of chaotic modulated time delay networks in presence of noise}
\author{Soumen Majhi\inst{1}, Bidesh K. Bera \inst{1}, Santo Banerjee \inst{2}\fnmsep\thanks{\email{santoban@gmail.com}}, \and Dibakar Ghosh \inst{1}\fnmsep\thanks{\email{dibakar@isical.ac.in}}}
\institute{\inst{1} Physics and Applied Mathematics Unit, Indian Statistical Institute, Kolkata-700108, India\\
\inst{2} Institute for Mathematical Research, Universiti Putra Malaysia, Selangor, Malaysia}
\abstract{
We study the constructive role of noises in a Lorenz system with functional delay. The effect of delay can change the dynamics of the system to a chaotic one from its steady state.  Induced synchronization with white and colored (red and green) noises are observed between two identical uncoupled systems and enhancement of synchrony is also observed with unidirectional coupling. We investigate both the phenomena in a globally coupled network in the presence of white and color noises. } 
\maketitle
\section{Introduction}
\label{intro}
Noise is omnipresent in natural and man-made systems. Noise plays a destructive and constructive role. Studies of coupled or uncoupled nonlinear systems driven by common forcing or environment have become an important issue during the last few decades \cite{dest_noise1,dest_noise2,indu_noise1,indu_noise2,indu_noise3,enh_noise1,enh_noise2,enh_noise3}. The destructive role of noise in coupled systems has been studied as a robustness test for synchronizations in the presence of noise because noise is always present in any experiment in nature \cite{dest_noise1,dest_noise2}. The constructive role of noise is noise-induced synchronizations \cite{indu_noise1,indu_noise2,indu_noise3} in uncoupled systems or the enhancement of synchrony \cite{enh_noise1,enh_noise2,enh_noise3} in weakly coupled systems. The first study of noise-induced synchronization was observed between a pair of uncoupled sensory neurons \cite{neuron}. Numerical and experimental studies on noise-induced synchronization have been conducted on physical systems such as lasers, electronics circuits and biological systems \cite{neuron,laser}. The effect of noise on synchronization is studied in low dimensional systems when they are coupled. Another constructive role of noise has been studied extensively in the context of stochastic resonance \cite{benzi,gamma,moss} and coherent resonance \cite{longtin,gang}. Most of the previous studies \cite{indu_noise1,indu_noise2,indu_noise3,enh_noise1,enh_noise2,enh_noise3} on noise-induced synchronization or noise enhanced synchrony have used white noise, which has infinite variance and no time correlation. But in many practical situations researchers considered colored noise in which time correlation is important. However, our interest here is in colored noise, namely green and red noise because they are negatively and positively correlated, respectively. The effect of colored noise has been discussed for fixed point \cite{uchida}, limit cycle \cite{devis}, and chaotic systems\cite{devis}. Little attention has been paid to modulated time delayed chaotic systems.
\par Time delay dynamical systems have two properties that make the study more interesting. Firstly, deterministic time delay dynamical systems have a nonzero memory i.e. they do not satisfy the Markov property \cite{markov}. Second, a nonlinear dynamical system with a fixed time delay $\tau $ gives rise to infinite dimensional dynamics on the phase space $C(-\tau ,0)$ of continuous functions on the interval $(-\tau, 0)$ \cite{hyper}. Thus they are associated with infinite dimensional chaotic attractors. For large values of time-delay, dynamical systems generated by a scalar delay differential equation become hyperchaotic, i.e. more than one positive Lyapunov exponent \cite{farmer}. So deterministic delay differential equations can have quite chaotic dynamics but it is not clear if a small noise will have any significant effect on such hyperchaotic evolution. Recently, dynamical systems with modulated time delay $\tau(t)$ have received more attention due to their potential application in secure communications \cite{ghosh07,ghosh08}. Previously most of the studies on noise-induced synchronizations \cite{indu_noise1,indu_noise2,indu_noise3} and noise-enhanced synchronizations \cite{enh_noise1,enh_noise2,enh_noise3} have been done for low dimensional systems.
In this paper, we extend the study of noise-induced and noise-enhanced synchronizations in modulated time-delay systems driven by a common noise. This study is important due to two simple facts. Since a delayed dynamical system has no Markovian property, the addition of noise or a stochastic perturbation does not change the dynamics to Markovian dynamics. Therefore well-known methods for Markovian process such as the Fokker-Planck equation cannot be used. Moreover it is not clear how small additions of white and colored noise can have any significant effect in infinite dimensional systems. The effect of noise, white as well as colored, on globally coupled modulated time-delay systems has not yet been studied.

\par The plan of the manuscript is as follows. In Sec. 2, we briefly discuss the modulated time-delay Lorenz system which we have considered for our study in synchronization. The noise-induced synchronization between two identical time-delay Lorenz systems driven by common white, red and green noises are discussed in Sec. 3. When two identical systems are coupled unidirectionally in the presence of common noise, an enhancement of synchronization occurs in the phase spaces of coupling strength and critical noise intensities, as illustrated in Sec. 4. In Sec. 5, we discuss the effect of white and colored noise in globally coupled systems. Finally, we summarize our results in Sec. 6.

\section{Modulated time delay Lorenz system}
We consider a Lorenz system with time delay \cite{ijbc_delay_lorenz} in the form
$$\dot x=\sigma(y-x)$$
$$\dot y=r~x-x~z(t-\tau)-y \;\;\;\; \eqno{(1)}$$
$$\dot z=x~y-bz(t-\tau)$$
where $\tau\geq 0$ is the time delay for the $z$ variable and $(\sigma, r, b)$ are system parameters. Without time delay, i.e. $\tau=0$ the system is chaotic for the set of parameter values $\sigma=10, r=28$ and $b=8/3.$ The fixed points of system (1) are $E_0=(0, 0, 0),$ $E_{1, 2}=(\pm \sqrt{b(r-1)}, \pm \sqrt{b(r-1)}, r-1)$. For $\sigma=10, b=8/3$ and $\tau=0.0,$ the zero fixed point $E_0$ is stable for $r\leq 1.0$ and the non-zero fixed points $E_{1, 2}$ are stable for $1<r<24.74.$ 
\begin{figure}
\resizebox{0.9\columnwidth}{!}{%
\includegraphics{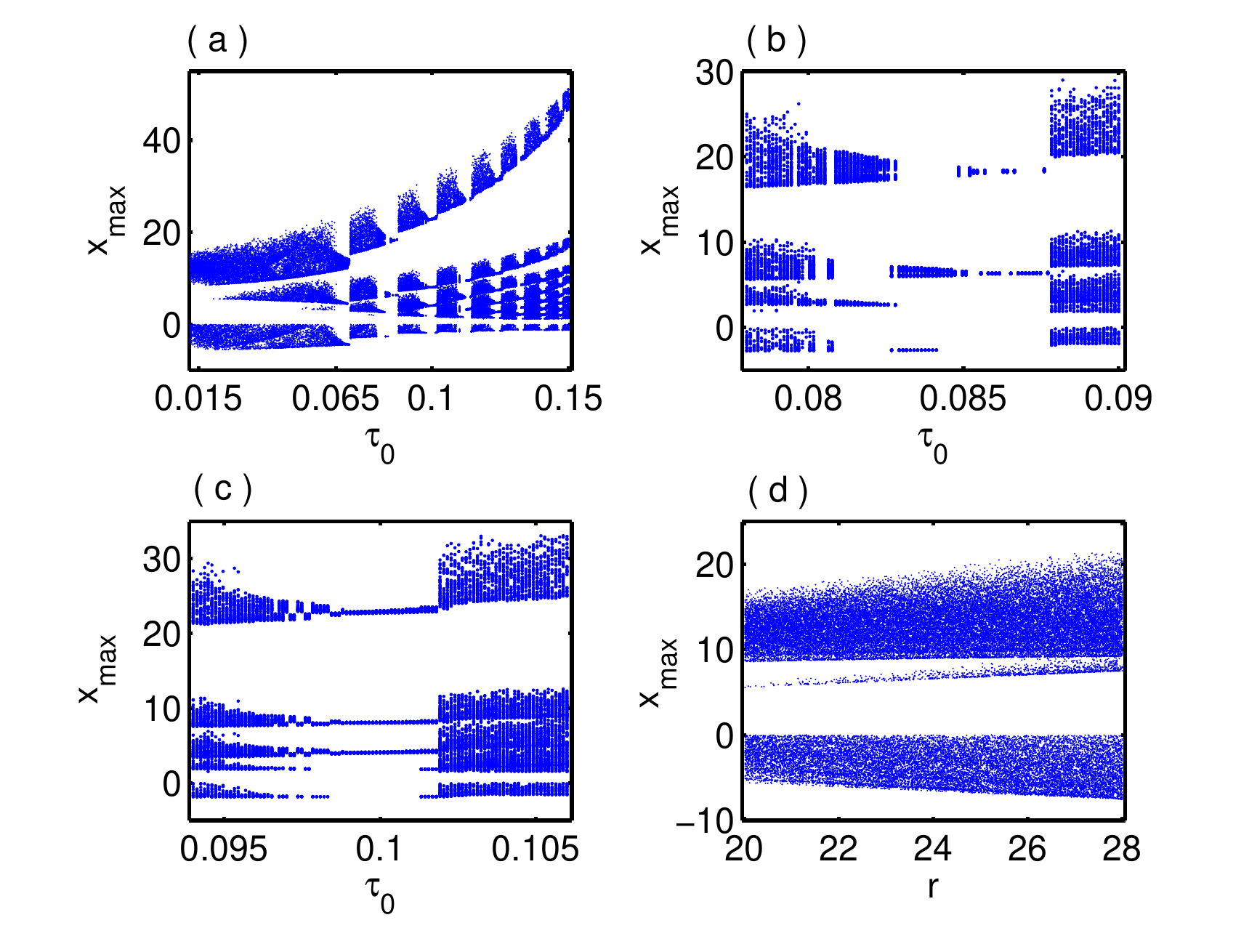} }
\caption{(color online) Bifurcation diagram of a modulated time-delay Lorenz system (1) by (a) varying zero-frequency component $\tau_0$ of $\tau(t)$ with $a=0.001, \omega=0.5, r=20$, (b) periodic window for $\tau_0 \in[0.078, 0.09]$, (c) for $\tau_0 \in[0.094, 0.106]$. (d) Bifurcation diagram by varying $r$ with $\tau_0=0.02, a=0.001$ and $\omega=0.5.$ The other parameters are fixed at $\sigma=10$ and $  b=8/3.$}
\label{fig:1}       
\end{figure}

\par Next we consider the time delay parameter $\tau$ as a function of time instead of a constant delay as $\tau(t)=\tau_0+a ~\mbox{sin}(\omega t)$, where $\tau_0$ is the zero-frequency component, $a$ is the amplitude and $\omega/{2 \pi}$ is the frequency of the modulation \cite{ghosh07,ghosh08}. For $\sigma=10, b=8/3 $ and $r=20$, the Lorenz system (1) without delay time has a stable steady states at $E_{1, 2}$ and it becomes chaotic in the presence of time delay modulation. The bifurcation diagram produced by varying the zero-frequency component $\tau_0$ is shown in Fig. 1(a) for $a=0.001, \omega=0.5$ and $r=20$. Fig. 1(a) shows that for $\tau_0$ lying between 0.078 to 0.105, many fascinating changes take places, i.e. periodic window and crises. Specially, Fig. 1(b) shows that there is a periodic window for [0.078, 0.09]. At $\tau_0=0.087$, a sudden destruction of the chaotic attractor occurs and it is replaced by a periodic one. With further increases of $\tau_0$, sudden widening or sudden increases in the size of the chaotic attractor occurs through crises. From Fig. 1(c) it is seen that a periodic window of period three occurs for $\tau_0\in[0.094, 0.106]$.  The system is always chaotic for all values of $r$ in $20\leq r\leq 28$ for $\tau_0=0.02, a=0.001$ and $\omega=0.5$ are shown in Fig. 1(d).
\begin{figure}
\resizebox{1.2\columnwidth}{!}{%
\includegraphics{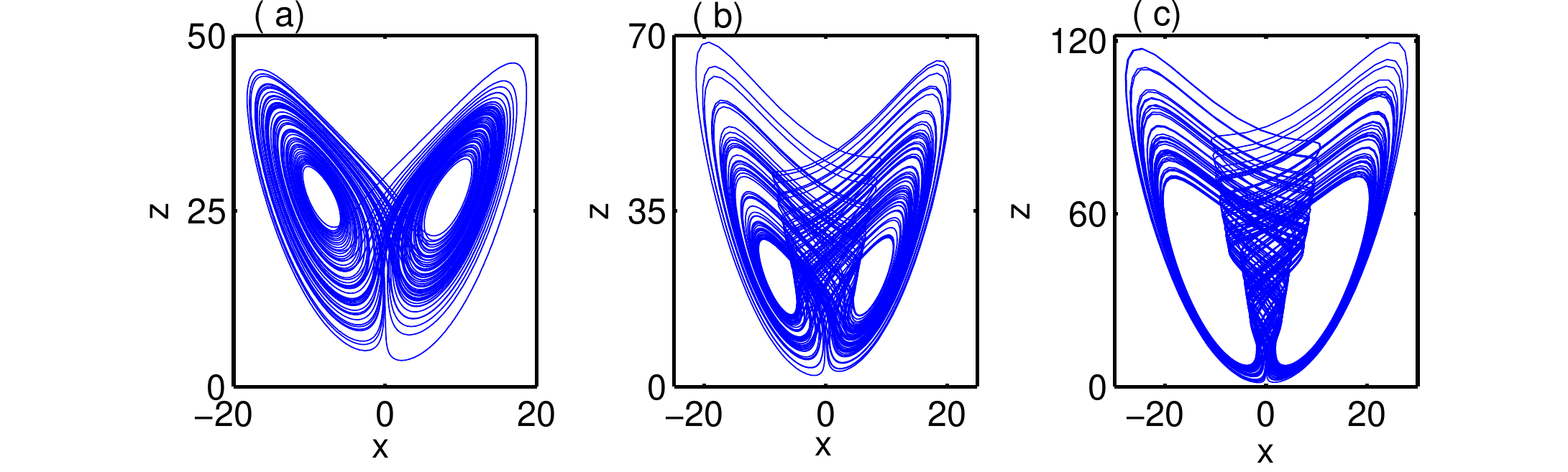} }
\caption{(color online) Chaotic attractor of (a) Lorenz system, $r=28.0, \tau=0.0$, (b) Chen attractor, $r=20.0, \tau_0=0.06$ and (c) Lu attractor, $r=20.0, \tau_0=0.09$. The other parameters are fixed at $\sigma=10, b=8/3, \omega=0.5$.}
\label{fig:1}       
\end{figure}

\par Then for different values of non-zero time delay $\tau(t)$, the delay Lorenz system (1) can exhibit not only the Lorenz system but also a Chen attractor and Lu attractor without changing the values of $r, \sigma$ and $b$. For $\tau=0.0,$ the model (1) becomes standard Lorenz model without a delay time, while for $\tau_0=0.06$ and $\tau_0=0.09$, the arractor become the Chen and Lu system's attractors for $r=20, a=0.001, \omega=0.5$. Numerical simulation results for three typical values of time delays $\tau(t)=0.0$, $\tau_0=0.06$ and $\tau_0=0.09$, corresponding to the Lorenz, Chen and Lu systems are shown in Figs. 2(a), 2(b) and 2(c) respectively. A comparison of three  chaotic attractors clearly show the difference between the Lorenz, Chen and Lu systems. It is observed from Figs. 2(b) and 2(c) that one more chaotic branch is emerging from the base chaotic attractor, which imply they are more complex than the Lorenz chaotic attractor.

\section{Noise-induced synchronization}
The general form of noise-induced synchronization \cite{diba_physica} is in the form
$$\dot X=F(X, X_\tau)+\xi \eqno{(2a)}$$
$$\dot Y=F(Y, Y_\tau)+\xi \eqno{(2b)}$$
where $X, Y \in R^m, \;\; X_\tau=X(t-\tau),\; Y_\tau=Y(t-\tau)$ and $\xi(t)$ is the common noise. Systems (2a) and (2b) are driven by the common noise $\xi(t)$ without any interaction between them. The noises driving the two systems are either white noise $w(t)$ or colored noise. For colored noise we consider red noise $r(t)$ and green noise $g(t)$.

\par If $\xi(t)=w(t)$, the systems 2(a) and 2(b) are driven by Gaussian white noise with the properties $\left \langle w(t) \right \rangle =0$ and $\left \langle w(t) w(t') \right \rangle =2 D_w \delta(t-t'),$ where $D_w>0$ is the noise intensity, $\delta$ is Dirac's delta function, and $\left \langle \cdot  \cdot  \cdot \right \rangle $ denotes averaging over the realizations of $w(t)$.
\par For red noise, we replace $\xi(t)$ by $r(t)$, which can be generated by the Langevin equation \cite{red_noise} and the dynamic evolution of $r(t)$ is given by
$$\dot r(t)=-\alpha_r r+\alpha_r w  \eqno{(3)}$$ 
where $\alpha_r$ is a positive constant. The red noise satisfies the following properties: $\left \langle r(t) \right \rangle =0$ and $\left \langle r(t) r(t') \right \rangle =D_r \alpha_r e^{-\alpha_r |t-t'|}$ where $D_r$ is the red noise intensity. We see that the autocorrelation function delays exponentially for the stochastic process $r(t).$
\par For green noise, we replace $\xi(t)$ by $g(t)$ in equation 2(a) and 2(b). The dynamic evolution of green noise is represented by the equation 
 $$\dot g(t)=-\alpha_g g-\dot w  \eqno{(4)}$$
where $\alpha_g$  physically represents the inverse correlation time and is a positive constant. The green noise $g(t)$ satisfies the following properties: $\left \langle g(t)  \right \rangle =0$ and $\left \langle g(t) g(t') \right \rangle =2 D_g \delta(t)-D_g \alpha_g e^{-\alpha_g |t-t'|}$ where $D_g$ is the green noise intensity. Physically, equation (4) represents a simple circuit system where a resistor, a capacitor, and a white noise voltage source are connected in a series. If the voltage fluctuates according to Gaussian white process, the current behavior will involve a time derivative of a Gaussian process. The stochastic process $\dot w$ represents the violet noise which usually occurs in electronic circuit systems.
\begin{figure}
\resizebox{0.7\columnwidth}{!}{%
\includegraphics{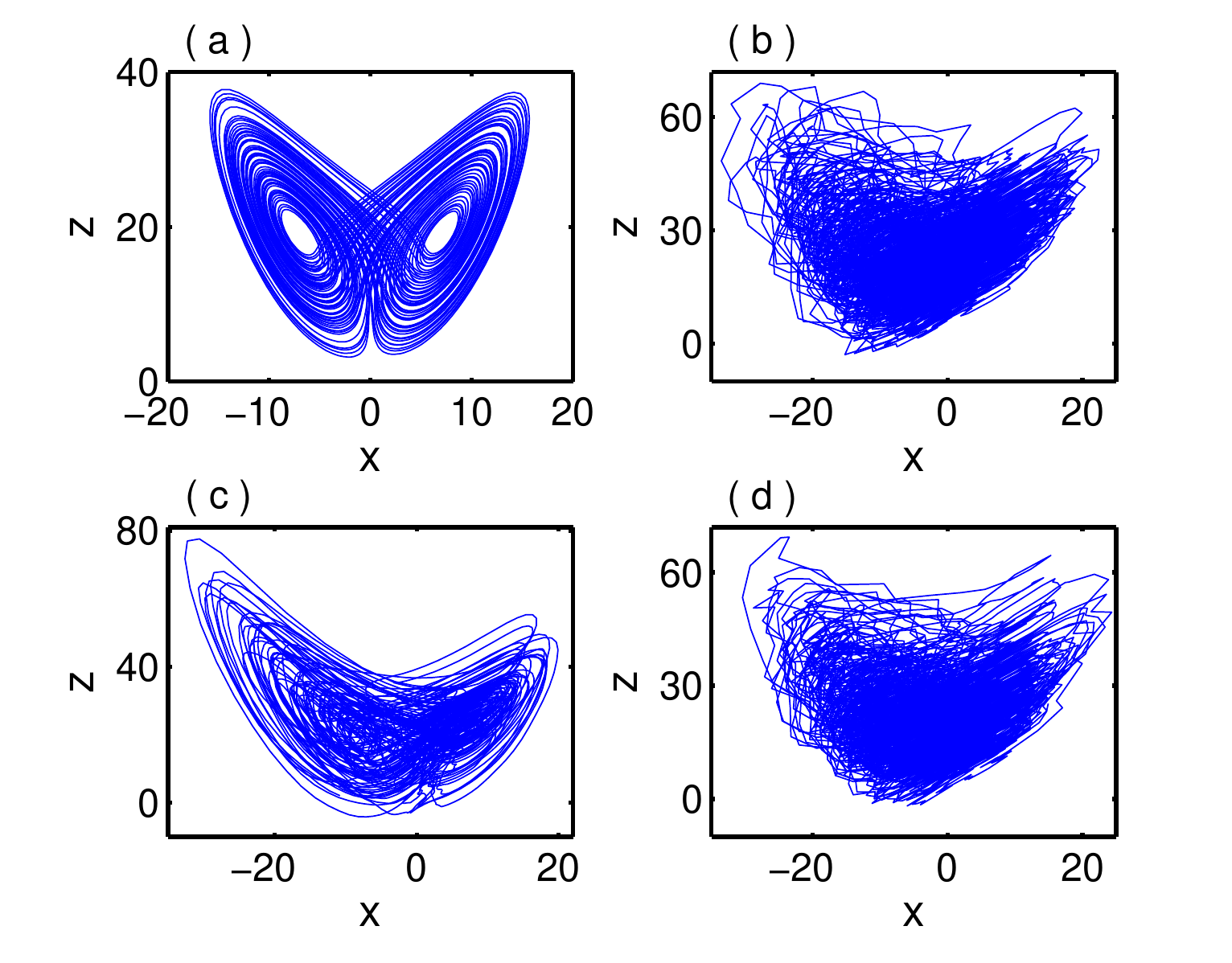} }
\caption{(color online) Chaotic attractors of a modulated time delay Lorenz system (1) in the presence of Gaussian white and color noise for different noise intensities: (a) without any noise, (b) Gaussian white noise with $D_w=20.0$, (c) red noise with $\alpha_r=3.0$, $D_r=52.0$ and (d) green noise with $\alpha_g=3.0$, $D_g=26.0$. The other parameter values are fixed at $\sigma=10, r=20, b=8/3, \tau_0=0.02, a=0.001$ and $\omega=0.5.$   }
\label{fig:1}       
\end{figure}

\par The system (1) is chaotic without any noise for the set of parameter values $\sigma=10, r=20, b=8/3, \tau_0=0.02, a=0.001,$ and $\omega=0.5$, the chaotic attractor is shown in Fig. 3(a). We consider $F(X, X_\tau)$ in equation 2(a) as $F(X, X_\tau)=(\sigma(y_1-x_1), r~x_1-x_1~z_1(t-\tau)-y_1, x_1~y_1-bz_1(t-\tau))^T$,  where $T$ represents the transpose.
The effects of white, red and green noise on an isolated Lorenz system (1) are shown in Figs. 3(b), 3(c) and 3(d) respectively for different noise intensities. 
\begin{figure}
\resizebox{0.95\columnwidth}{!}{%
\includegraphics{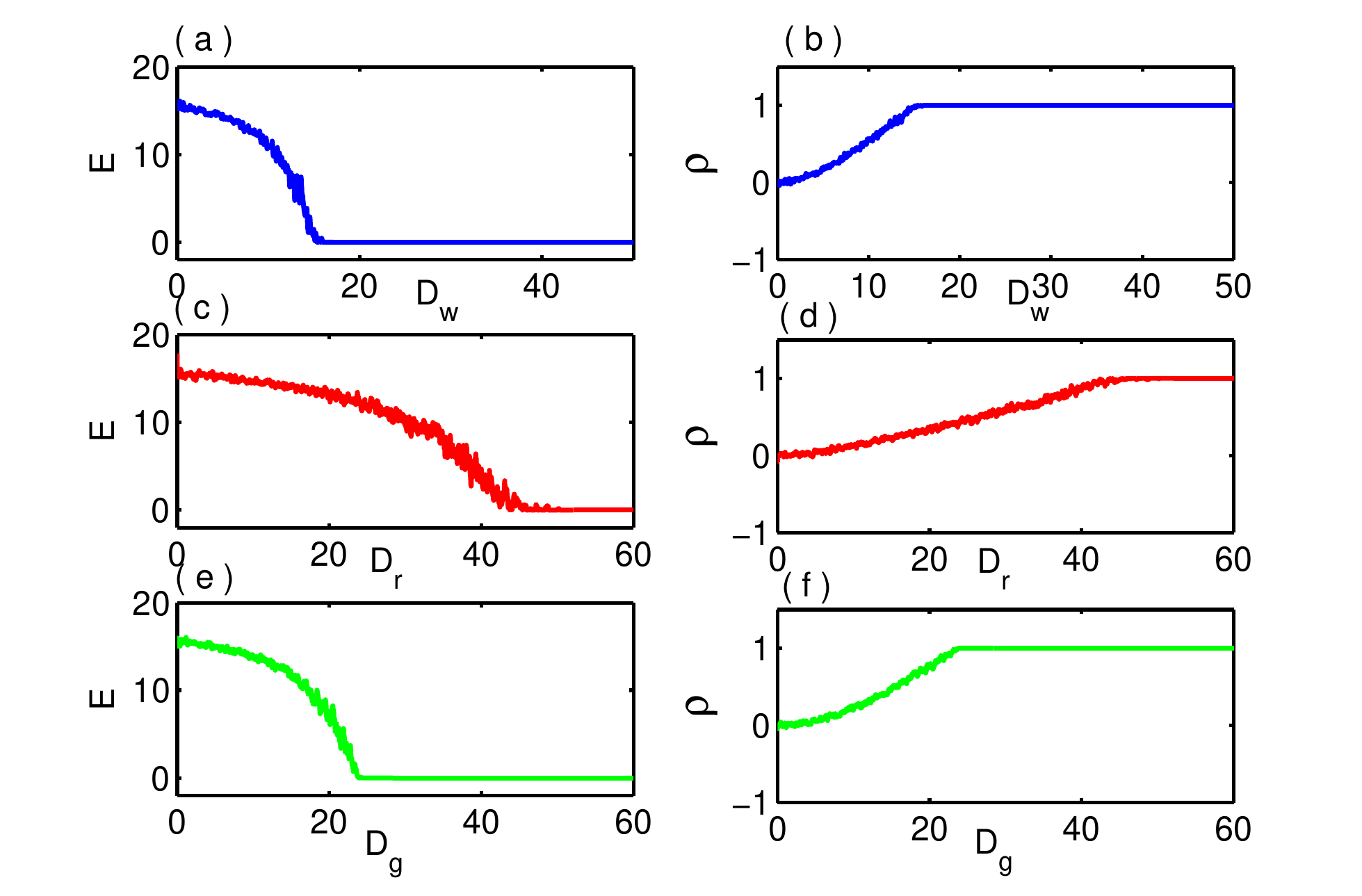} }
\caption{(color online) Variation of synchronization error $E$ and cross-correlation function $\rho$ for different values of noise intensities: (a, b) Gaussian white noise, (c, d) red noise and (e, f) green noise.}
\label{fig:1}       
\end{figure}

\par Complete synchronization is established between two systems 2(a) and 2(b) driven by common noise and without any coupling between them. This complete synchronization is known as noise-induced synchronization \cite{indu_noise1,indu_noise2,indu_noise3,senthil}. To quantify the noise-induced synchronization between 2(a) and 2(b), we define the complete synchronization error as 
$$E=\left \langle \sqrt{(x_2-x_1)^2+(y_2-y_1)^2+(z_2-z_1)^2} ~~\right \rangle,  \eqno{(3)}$$
where $\left \langle \cdot \cdot \cdot \right \rangle $ represents the time average. The synchronization error $E$ between systems 2(a) and 2(b) as a function of Gaussian white noise intensity $D_w$ is shown in Fig. 4(a). It is seen from this figure that the synchronization error $E$ decreases on increasing the Gaussian white noise intensity $D_w$ and for $D_w>15.9$, the synchronization error $E$ approaches zero which means the systems 2(a) and 2(b) are completely synchronized due to common noise. In order to better characterize the noise-induced synchronization, we define the cross-correlation function $\rho$ between two variables $x_1(t)$ and $x_2(t)$ as 
$$\rho=\frac{\left \langle x_1(t) x_2(t) \right \rangle }{\sqrt{\left \langle x_1^2(t) \right \rangle  \left \langle x_2^2(t) \right \rangle }} \eqno{(4)}$$ 
Fig. 4(b) shows the variation of cross-correlation function $\rho$ for different noise intensities $D_w$. The critical noise intensity is $D_w=15.9$, where the synchronization error is zero (from Fig. 4(a)) and the value of $\rho$ is near to unity (Fig. 4(b)) signify that the two systems 2(a) and 2(b) are completely synchronized with each other due to common white noise. Figs. 4(c) and 4(d) show the variation of the synchronization error $E$ and the cross-correlation function $\rho$ for different red noise intensities $D_r$ for $\alpha_g=3.0.$ From this figure we see that the critical red noise intensity for synchronization is $D_r=45$ which is significantly different from the critical white noise intensity. The variation of synchronization error $E$ and cross-correlation function $\rho$ for different values of green noise intensities $D_g$ are shown in Figs. 4(e) and 4(f) respectively. The critical green noise intensity for synchronization is $D_g=25.$  
\begin{figure}
\resizebox{0.5\columnwidth}{!}
{\includegraphics{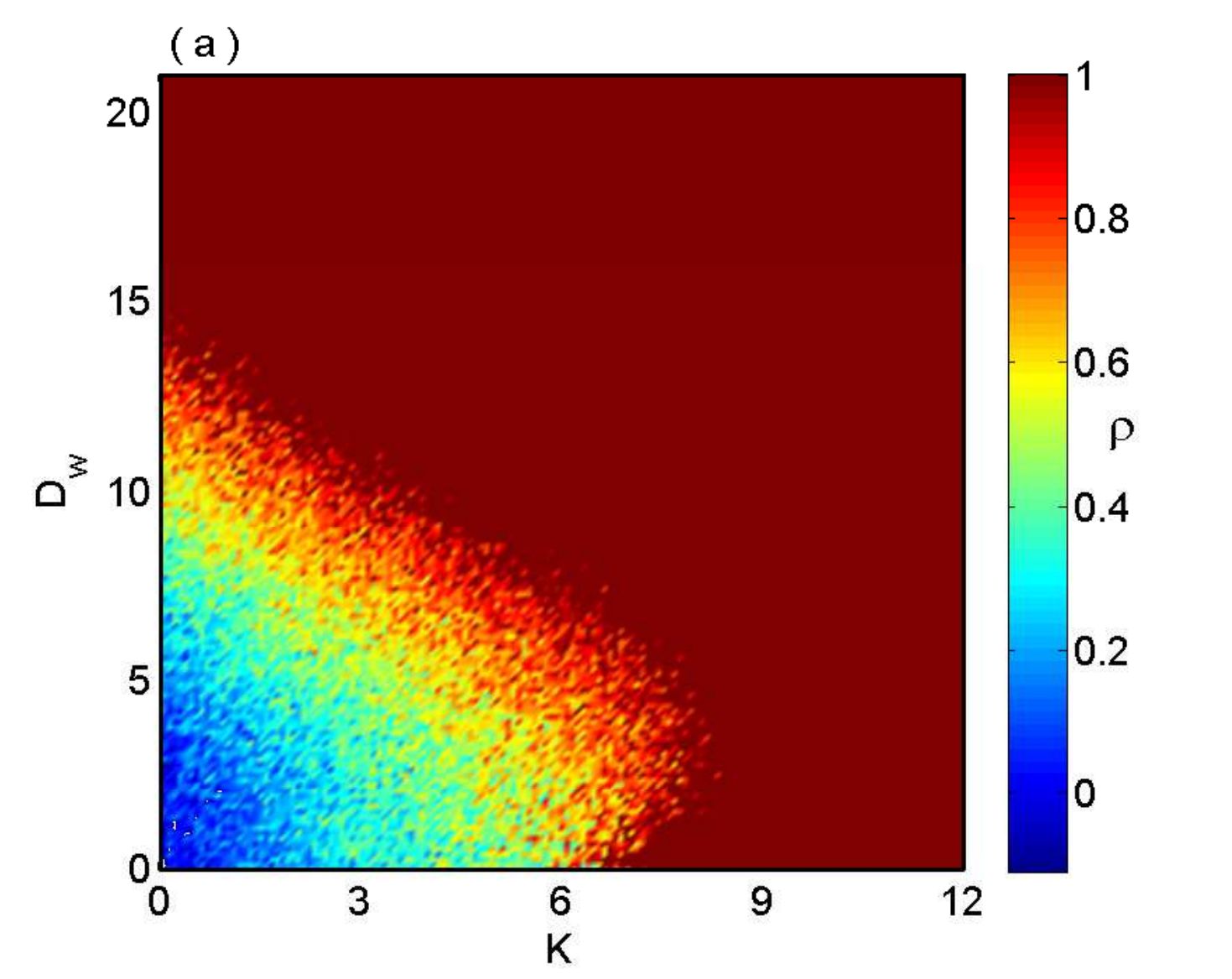} }
\resizebox{0.5\columnwidth}{!}
{\includegraphics{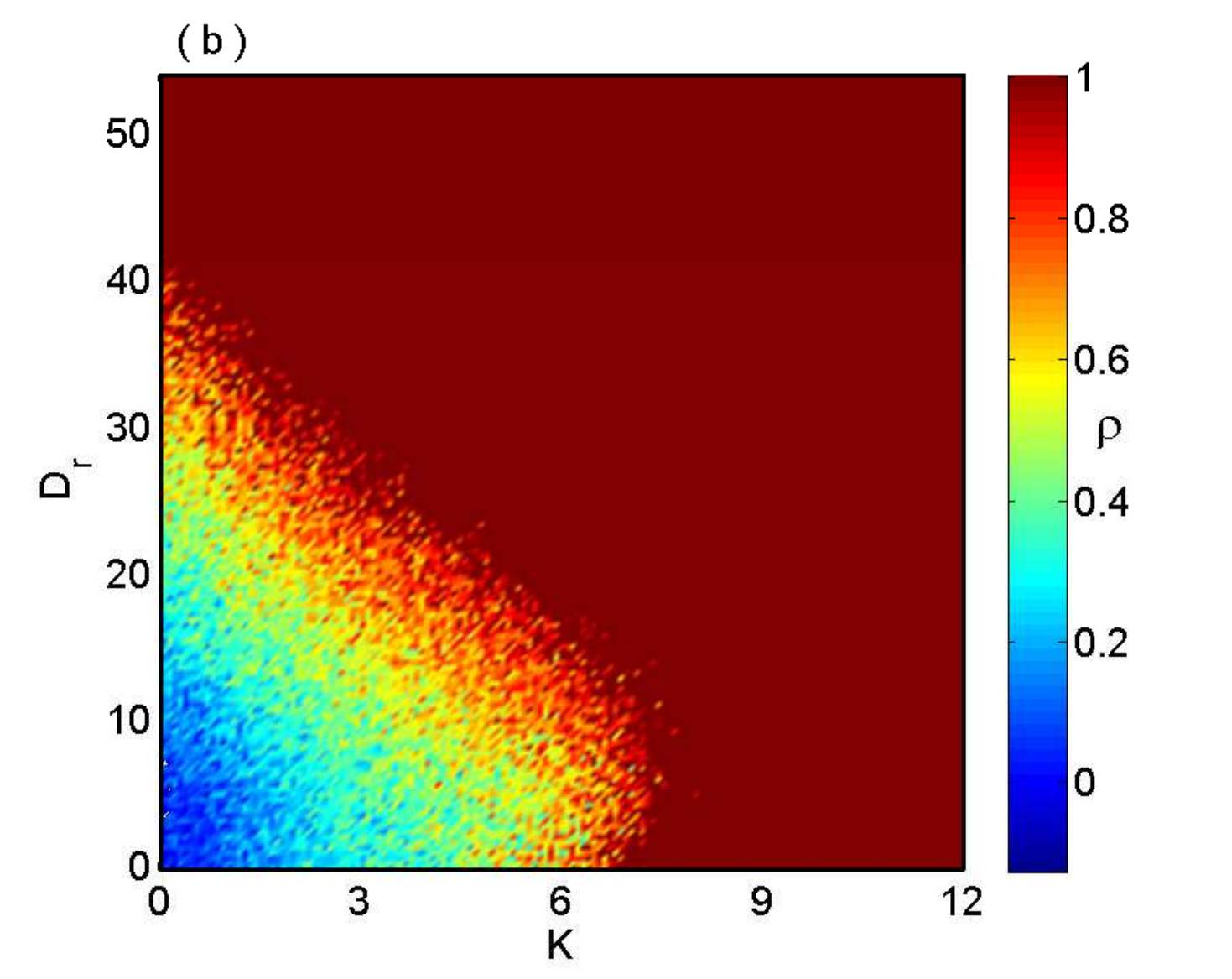} }
\resizebox{0.5\columnwidth}{!}
{\includegraphics{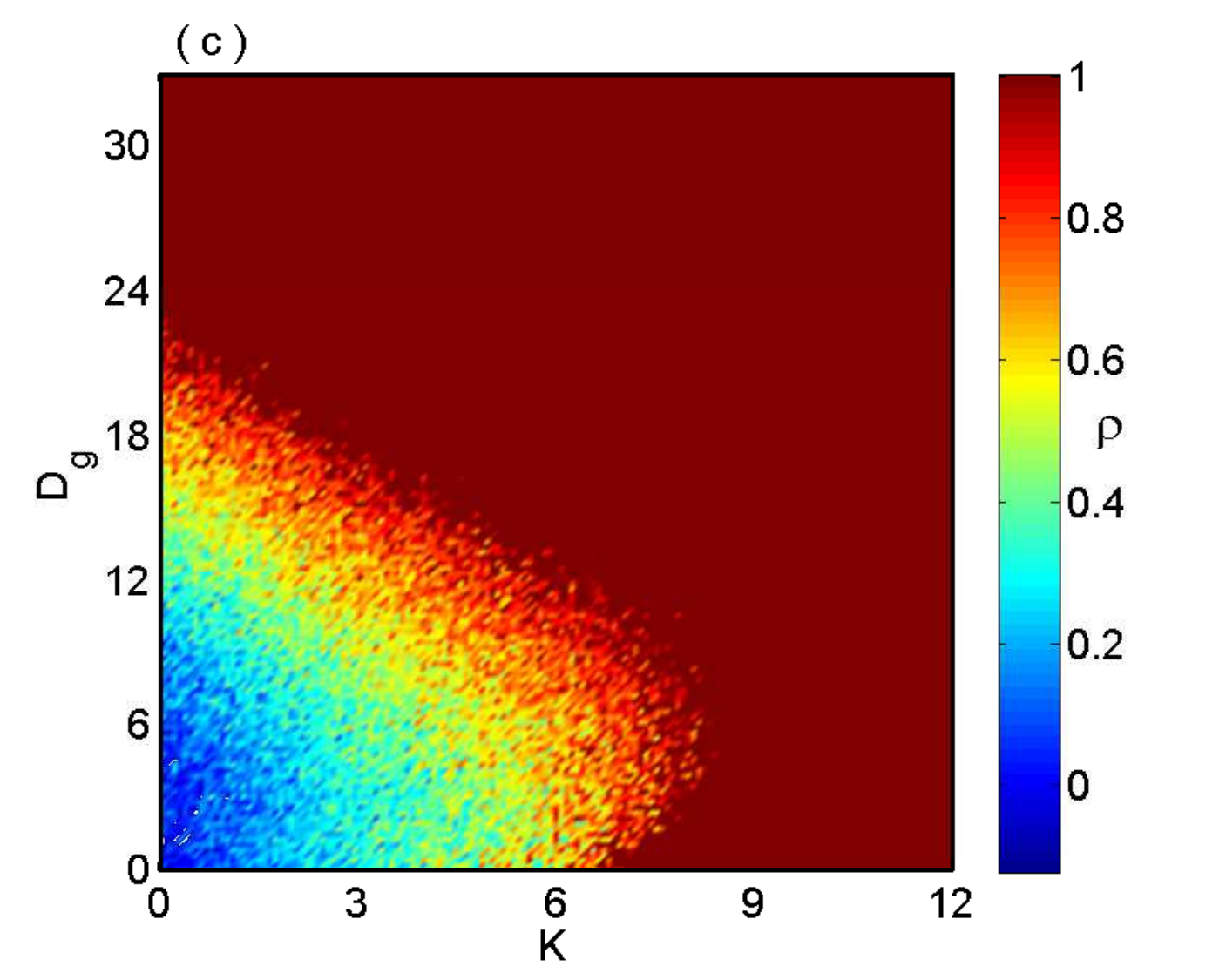} }
 \caption{(color online) Synchronization region in the parameter space of  the coupling strength and critical noise intensity: (a) Gaussian white noise, (b) red noise and (c) green noise. The other parameter values are fixed at $\sigma=10, r=20, b=8/3, \tau_0=0.02, a=0.001$ and $\omega=0.5.$}
\label{epsilon_12}
 \end{figure}

\section{Noise-enhanced synchronization}
We consider two identical delay Lorenz systems coupled unidirectionally in the presence of Gaussian white, red and green noise. The coupled system is as follows: 
$$\dot x_1=\sigma(y_1-x_1)+\xi(t)$$
$$\dot y_1=r~x_1-x_1~z_1(t-\tau)-y_1+\xi(t) \;\;\;\; \eqno{(5)}$$
$$\dot z_1=x_1~y_1-bz_1(t-\tau)+\xi(t)$$
and
$$\dot x_2=\sigma(y_2-x_2)+k(x_1-x_2)+\xi(t)$$
$$\dot y_2=r~x_2-x_2~z_2(t-\tau)-y_2+\xi(t) \;\;\;\; \eqno{(6)}$$
$$\dot z_2=x_2~y_2-bz_2(t-\tau)+\xi(t)$$
where $k\geq 0$ is the coupling strength. For $k=0,$ noise-induced synchronization occurred using white, red and green color noise, which are discussed in the previous section. The two-parameter phase diagram in the $(k - D_w)$ plane is shown in Fig. 5(a) where the cross-correlation function has been used as synchronization measure. From this figure it is clear that by increasing the Gaussian noise intensity $D_w$, two systems (5) and (6) are synchronized at a lower value of critical coupling strength $k$. The enhancement of synchronization occurs due to the presence of common noise on both the systems. Similarly, in the presence of red and green noise, the enhancement of synchronization also occurs and is shown in Figs. 5(b) and 5(c) respectively.  

 \begin{figure}
\resizebox{1.1\columnwidth}{!}{%
\includegraphics{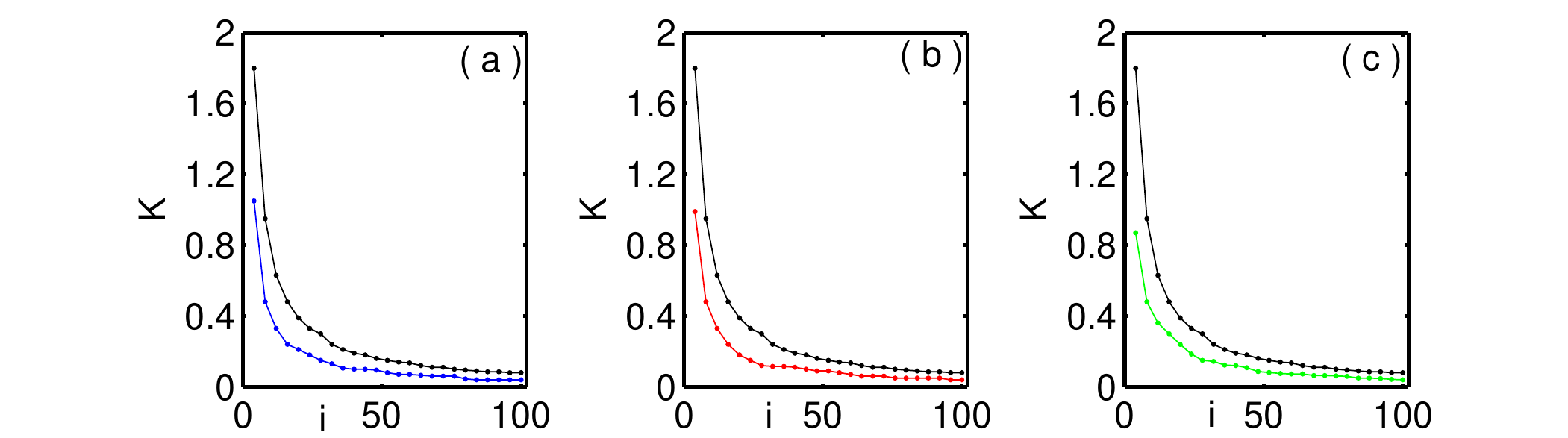} }
\caption{(color online) Effect of noise on synchronization in the network (7) with the number of oscillators: (a) Gaussian white noise, $D_w=10$, (b) red noise, $D_r=25$, and (c) green noise, $D_g=16$. The color lines are as follows: black = without noise, blue = in the presence of white noise, red = red noise and green = green noise. }
\label{fig:1}       
\end{figure}
\begin{figure}
\resizebox{0.5\columnwidth}{!}
{\includegraphics{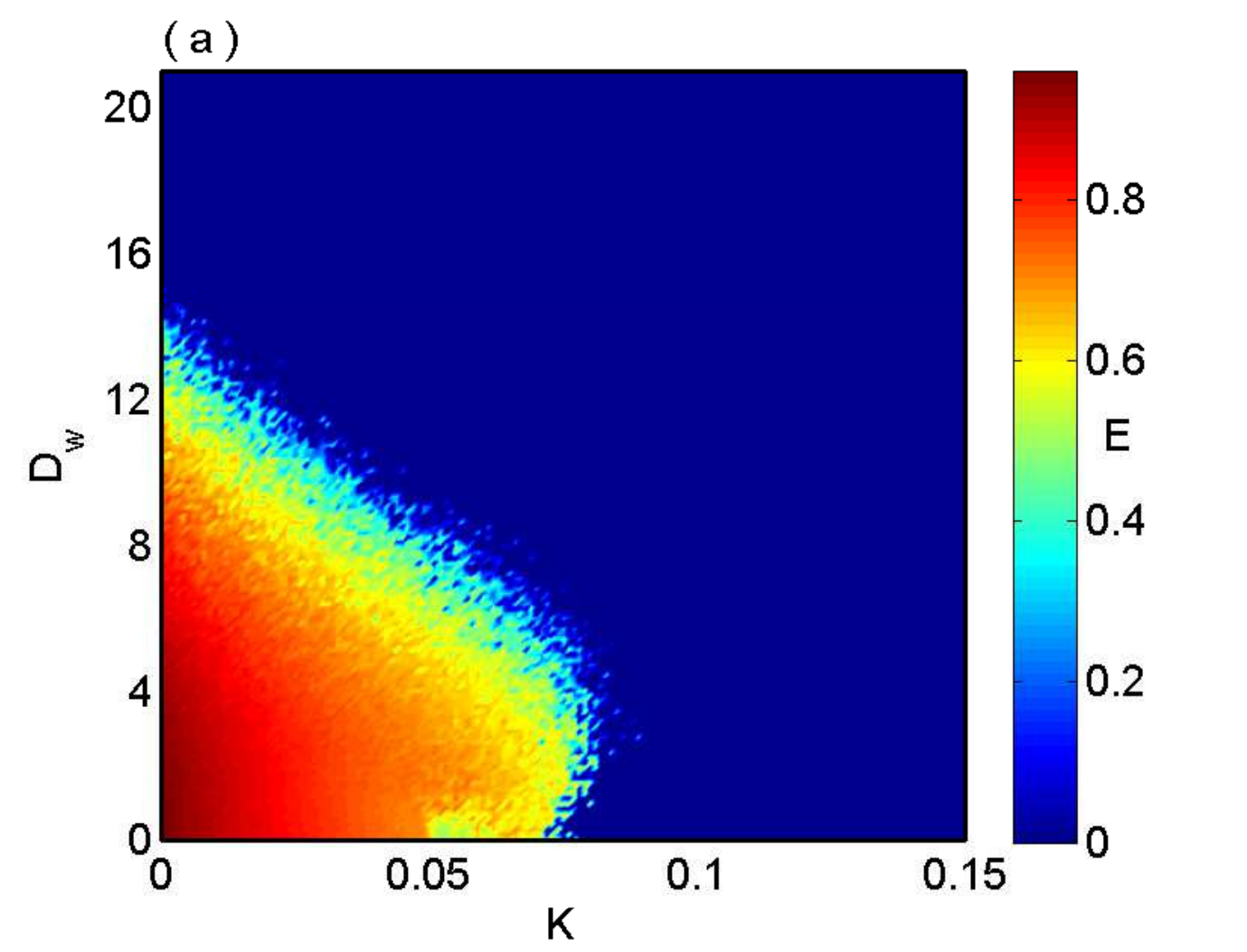} }
\resizebox{0.5\columnwidth}{!}
{\includegraphics{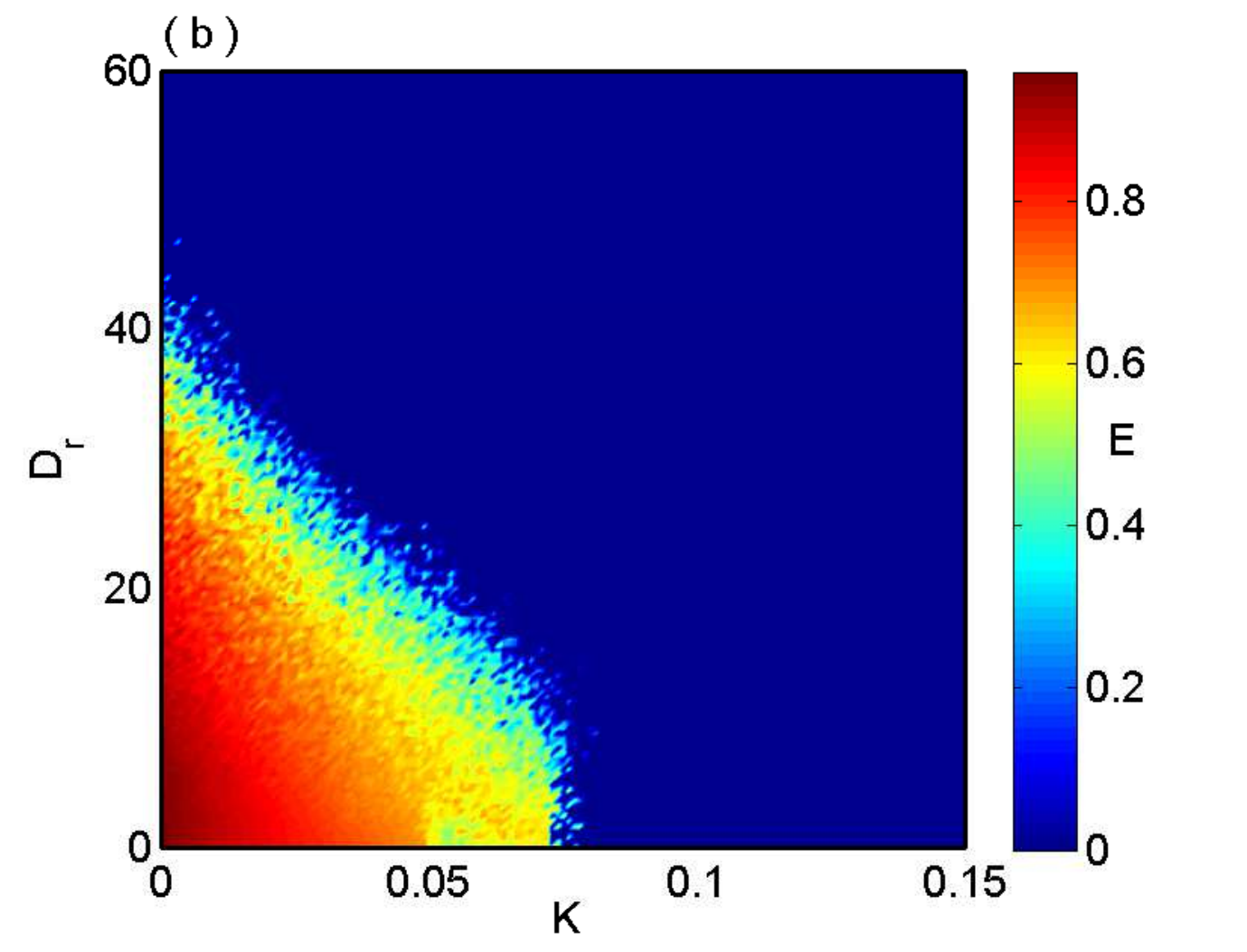} }
\resizebox{0.5\columnwidth}{!}
{\includegraphics{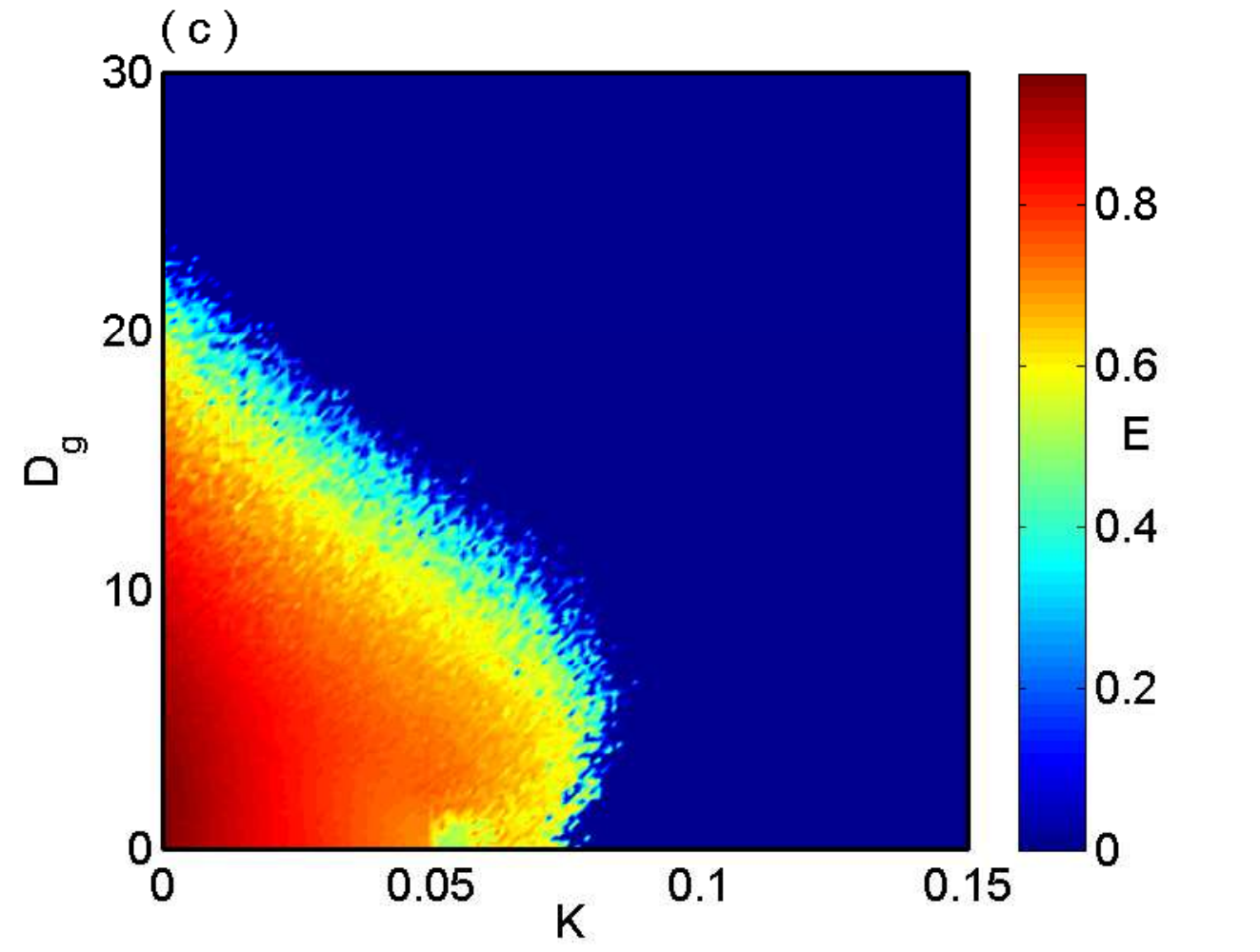} }
 \caption{(color online) Enhancement of synchronization in network of $N=100$ oscillators in the parameter space of  the coupling strength $k$ and critical noise intensities: (a) Gaussian white noise, (b) red noise and (c) green noise. The other parameter values are fixed at $\sigma=10, r=20, b=8/3, \tau_0=0.02, a=0.001$ and $\omega=0.5.$}
\label{epsilon_12}
 \end{figure}

\section{Network of globally coupled oscillators}
Next, we consider a network of globally coupled oscillators with additive noise and the equation is expressed as 
$$\dot x_i=\sigma(y_i-x_i)+k \sum_{j=1, i \not =j}^N (x_j-x_i)+\xi(t)$$
$$\dot y_i=r~x_i-x_i~z_i(t-\tau)-y_i+\xi(t) \;\;\;\; \eqno{(7)}$$
$$\dot z_i=x_i~y_i-bz_i(t-\tau)+\xi(t)$$
where $i=1, 2, \cdot \cdot \cdot ,N$ and $N$ is the number of oscillators in the network. Here $k$ is the coupling strength. Now we will discuss the collective dynamics of the globally coupled oscillators in the presence of three different types of noise, namely Gaussian white, red and green noise. The critical coupling strength for global synchronization by varying the number of oscillators is shown in Fig. 6 with and without any additive noise. In the absence of any noise, the critical coupling strength decreases by increasing the number of oscillators in the network as shown by black color lines in Fig. 6. But in the presence of white noise with noise intensity $D_w=10.0$, global synchronization occurred at lower critical coupling compared to the noise-free case (Fig. 6(a)) by blue color line.  Similarly, the critical coupling strengths in the presence of red and green noises are shown by red and green color lines in Figs. 6(b) and 6(c) respectively. We calculated the critical coupling strength by calculating the global synchronization error which is defined as, 
$$E=\left \langle \sqrt{\sum_{i=1}^{N-1} (x_{i+1}-x_i)^2} ~~\right \rangle  \;\;\;\; \eqno{(8)}$$

\par We have also estimated the global synchronization error $E$ by changing the noise intensities and coupling strength to check whether noise can enhance global synchronization for a wide range of noise intensities and coupling strength. We take the number of oscillators $N=100$. A two parameter $(k-D_w)$ synchronization region for the range $k \in (0, 0.15)$ and $D_w \in (0, 22)$ is shown in Fig. 7(a). From this figure it is shown that by increasing the noise intensities $D_w$, global synchronization occurred at lower values of coupling strength $k$. Similarly, the synchronization region in the presence of red and green noise are shown in Figs. 7(b) and 7(c) respectively.

\section{Summary}
In this paper, we discussed the constructive role of white and colored noise in a modulated time-delay Lorenz system. Color noise exists in many physical systems such as electronic circuit systems where the time correlation is important. For colored noise, we consider red and green noises which are positively and negatively correlated. Noise-induced synchronization is observed between two identical uncoupled modulated time-delay Lorenz systems and an enhancement in synchrony is observed when they are coupled unidirectionally. We observed both the phenomena in a globally coupled network of modulated time delay systems in the presence of white and color noises.
\par The present work is relevant for applications in many disciplines. For example, in neuroscience, a group of uncoupled sensory neurons are synchronized through common external noise \cite{neuron}.  In ecology, it is known that, due to common climate fluctuations, populations of plants exhibit large-scale synchronized flowering and production of seed crops also fluctuate synchronously from year to year  \cite{ecology}. One could examine the noise-induced synchronization using electronic circuits such as those studied experimentally in Ref. \cite{circuit_noise,mixed}

\end{document}